# The unique chemical reactivity of a graphene nanoribbon's zigzag edge

De-en Jiang[1,*], Bobby G. Sumpter[2,3], Sheng Dai[1]

[1]Chemical Sciences Division and [2]Computer Science and Mathematics Division and [3]Center for Nanophase Materials Sciences, Oak Ridge National Laboratory, Oak Ridge, Tennessee 37831

The zigzag edge of a graphene nanoribbon possesses a unique electronic state that is near the Fermi level and localized at the edge carbon atoms. We investigate the chemical reactivity of these zigzag edge sites by examining their reaction energetics with common radicals from first principles. A "partial radical" concept for the edge carbon atoms is introduced to characterize their chemical reactivity, and the validity of this concept is verified by comparing the dissociation energies of edge-radical bonds with similar bonds in molecules. In addition, the uniqueness of the zigzag-edged graphene nanoribbon is further demonstrated by comparing it with other forms of $sp^2$ carbons, including a graphene sheet, nanotubes, and an armchair-edged graphene nanoribbon.

*To whom correspondence should be addressed. E-mail: jiangd@ornl.gov. Phone: (865)574-5199. Fax: (865) 576-5235.

## I. INTRODUCTION

From graphite to fullerenes to nanotubes, carbon displays the versatility of $sp^2$ hybridization coupled with the right valence. The ability of carbon atoms to adopt $sp^2$ hybridization dictates many structural and physical properties of carbon materials, for example, the stability of a graphene sheet. Recent advances in graphene-based materials[1,2] exemplify the many possibilities brought about by the simple hexagonal structure of a graphene sheet.

The importance of an edge to a graphene sheet parallels that of a surface to a crystal. Cutting through an infinite graphene sheet [Fig. 1(a)], one first breaks C-C σ bonds and then obtains two semi-infinite graphene sheets, each with a one-dimensional edge. The dangling σ bonds at the edges can be saturated with hydrogen (so-called hydrogenated or hydrogen-terminated edges) and all the carbon atoms remain $sp^2$ hybridized. Depending on the cutting direction, two unique types of edges can be obtained: zigzag [Fig. 1(b)] and armchair [Fig. 1(c)]. The cutting also introduces a boundary at the edge to the previously fully delocalized π-electron system. It turns out that the geometry of the edge makes a huge difference in the π-electron structure at the edge. By constructing an analytical solution to the edge state, Nakada et al.[3] showed that the zigzag edge in a semi-infinite graphene sheet gives rise to a degenerate flat band near the Fermi level for the k vector between $2\pi/3$ and π. For k=π, the wavefunction is completely localized at the edge sites, leading to a so-called localized state at the zigzag edge. This flat-band feature and its corresponding localized state are unique to the zigzag edge (they are completely absent from the armchair edge).

Using the DV-Xα method with the linear combination of atomic orbitals (LCAO) bases, Kobayashi[4] in 1993 predicted the existence of such a flat band and localized state on a zigzag-edged vicinal graphite surface. Independently, Klein[5] analytically examined the band structure of the simple Hückel model for several graphene ribbons with zigzag edges, but the flat bands were predicted to be in $0 < k < 2\pi/3$ with a small gap at the Fermi level. This description was subsequently modified by the same author.[6] More thorough theoretical investigations were presented by Fujita, Nakada, and others.[3,7-16] Encouragingly, the zigzag edge was recently observed by scanning tunneling microscopy on highly oriented pyrolytic graphite (HOPG) surfaces and the localized state at the zigzag edge was confirmed by scanning tunneling spectroscopy (STS).[17-19]

The semi-infinite graphene sheet can be cut again parallel to the edge, to generate a graphene ribbon with two edges (Fig. 2). If the ribbon width is within the nanometer range, the effect of edge atoms is more pronounced[3] and the two edges can interact with each other. Theoretical studies using the Hubbard model with the unrestricted Hartree-Fock approximation revealed weak ferromagnetism along one edge and antiferromagnetism between two edges (one edge spin-up, the other spin-down) on a zigzag-edged graphene nanoribbon (ZGNR).[7] Here the remarkable thing is that the magnetism arises from a system made of only $sp^2$ carbon *without σ-dangling bonds*. The magnetism of ZGNR has been used to explain observed magnetic properties in nanographite materials.[20,21] Very recently, Son et al. predicted half metallicity in ZGNRs from first principles.[22] The half metallicity is caused by opposite responses of energy bands to the external electric field for the up and down spins.

Although H-free carbene-like zigzag edges[23] and H-free dangling σ-bond zigzag edges[24] have also been proposed to explain magnetism in carbon materials, it seems unlikely that those edge sites can survive under ambient conditions (room temperature in air) due to their high chemical reactivity. The zigzag state observed by STS in air at room temperature has been attributed to zigzag edges with saturated σ-bonds.[19]

To date, almost all theoretical and computational studies of graphene zigzag edges have focused on the physical aspects, such as electronic structures and magnetic properties. We believe that the localized state at the zigzag edge should also offer interesting chemical properties. Due to the nonbonding character of the localized state and the closeness of the flat band to the Fermi level, the zigzag edge sites should resemble a radical. But how the ZGNR distributes its π-electrons (Sec. III.A) and how these π-electrons respond to external chemical stimuli (Sec. III.B) are what we want to know from first-principles calculations. To show that ZGNRs are indeed unique, we compare ZGNRs with a graphene sheet, nanotubes, and an armchair edge, for their reactivity with atomic hydrogen (Sec. III.C). We discuss the implications of our work in Sec. III.D.

**II. COMPUTATIONAL METHODS**

Using the Vienna Ab Initio Simulation Package (VASP),[25,26] we performed density functional theory (DFT) calculations with a planewave bases and periodic boundary conditions and within the generalized-gradient approximation (GGA) for electron exchange and correlation.[27] The projector-augmented wave (PAW) method[28,29] within the frozen core approximation was used to describe the electron-core interaction. A slab model was used for the graphene sheet with an optimized C-C bond length (1.426 Å) and a 16-Å vacuum layer between the sheets. To model the graphene ribbons, a 10-Å vacuum layer was used to separate two neighboring edges. All atoms in the unit cell were allowed to relax and the force tolerance was set to 0.025 eV/Å. A kinetic energy cutoff (450 eV) was used, and Monkhorst-Pack k-meshes were employed to sample the Brillouin zone. Uncertainty in the interaction energy between a radical and a graphene nanoribbon (GNR), due to the kinetic energy cutoff and the k-point sampling, was estimated to be ~0.02 eV. Denser k-meshes were used to obtain the electronic density of states. The local magnetic moment (that is, net spin polarization on an atom) was obtained by integrating the local density of states up to the Fermi level for spin-up and spin-down states separately and then taking the difference between the two.

**III. RESULTS AND DISCUSSION**

**A. Magnetic phases of ZGNRs.**

Although various theoretical and computational methods have been employed to examine the magnetic properties of ZGNRs,[7,22,24] comparison of stability of different magnetic phases has not been fully reported for the hydrogen-terminated ZGNRs. So we first quantify the energetic differences among the magnetic phases of ZGNRs to find the ground state, and then examine its electronic and magnetic properties, as a preparation for the subsequent discussion of the chemical reactivity.

Fig. 2 shows the various ZGNR models used in the present study. All the carbon atoms are in their $sp^2$ hybridization state and the dangling bonds are saturated with hydrogen atoms. The width of the ribbon is indexed by a number, N, listed on the left of the three models. Three widths that are close to 1 nm were examined and the stability of the three magnetic phases is given in Table I. The antiferromagnetic (AFM) phase is the most stable, followed by the ferromagnetic (FM) phase,[30] and then the nonmagnetic (NM) phase. The difference between the AFM and FM phases is smaller than that between the FM and NM phases, and changes only slightly from N=4 to N=6. We also computed the total magnetic moment for the FM phase, and found that it increases slightly from N=4 to N=6 (Table II). This is also the case for the local magnetic moment of the edge carbon atoms that are connected to the hydrogen atoms (subsequently we call these carbon atoms edge carbons or edge sites). In the FM phase, these edge carbon atoms have much larger local magnetic moments than the non-edge carbon atoms (Fig. 3). In terms of magnitude, the AFM phase gives the same trend for the local magnetic moments as the FM phase, but the two edges have opposite signs (Fig. 3). Another point is that the local magnetic moment of the edge carbon atoms is larger for the AFM phase than the FM phase. This larger value of the AFM phase may be related to its lower energy, because larger magnetic moments indicate stronger exchange interaction.

The stability of the AFM phase for the ZGNRs can be understood with a Hubbard model, as shown by Fujita et al.[7] They predicted a local magnetic moment of 0.19 for the zigzag edge carbon atoms (N=10, U/t=0.1),[31] in agreement with our results. Son et al.[22] examined ZGNRs with DFT in the local spin density approximation and Lee et al.[24] investigated ZGNRs without hydrogen termination using DFT-GGA. The stability trend from both studies is the same as the one we have found here.

The appearance of a flat band near the Fermi level due to the edge state has been shown in various investigations of the band structure of ZGNRs.[3,4,7] This flat band, due to a localized state at the edges, results in a sharp peak near the Fermi level in the local density of states (LDOS) for the edge carbon atoms (Fig. 4). For the NM phase, the Fermi level bisects the sharp peak, which leads to instability and is subject to Stoner magnetism.[32] As a result, the two spin states shift in the opposite direction relative to the Fermi level. The finite DOS at the Fermi level for the FM phase indicates that the phase is metallic, whereas a small gap opens up for the AFM phase, indicating that the phase is semiconducting. Son et al.

have showed that the band gap in the AFM phase decreases to zero for electrons of one spin and increases for electrons of the other spin when a strong enough external electric field is applied across the ZGNRs, leading to so-called half-metallicity.[22]

Although direct evidence of magnetism in either semi-infinite zigzag graphene edge or ZGNRs has not been reported, some researchers have connected theoretical magnetism of ZGNRs with measured magnetism in nanographite materials.[20,21] Detailed discussion of magnetism in carbon materials is beyond the scope of the present paper.

**B. The chemical reaction between ZGNRs and common radicals.**

We have shown that spin-polarized π-electrons are localized on the zigzag carbon atoms, which from a chemical view, prompts us to think of them as a "partial radical". That is, these ZGNRs have unpaired π-electrons distributed mainly on the two edges, but on average each edge carbon atom has only 0.14 electrons. Due to the partial radical character, these edge carbon atoms should offer special chemical reactivity, comparing with non-edge ribbon carbon atoms, armchair carbon atoms, or nanotube carbon atoms, which shows little or no radical character. We will show the chemical reactivity of ZGNRs with common radical in this section and compare them with other $sp^2$ carbons in the next.

First, we examine the reaction of the zigzag edge with a hydrogen atom, the simplest radical. The bond dissociation energy (BDE)[33] is calculated for the newly formed C-H bond in which C is an edge carbon. One can see that the BDE changes only slightly for widths of N=4 to N=6 (Table III). Comparing with other $C(sp^3)$-H BDEs (4.553 eV for $C_2H_5$-H and 4.315 eV for cyclo-$C_6H_{11}$-H),[34] the C-H bond formed at the zigzag edge has a strength of ~60% of the C-H bond between a molecular carbon radical and H. This observation indicates that a "partial radical" concept is useful to characterize the chemical reactivity of the zigzag edge. Of course, a question arises accordingly: why the edge carbon has a partial charge of ~0.14 e but the edge C-H bond has a strength of ~60% that of a common C-H bond? This question brings up another aspect of the chemical reactivity at the zigzag edge. Although the localized state at the zigzag edge offers only a partial amount of the π-electron density on a per edge carbon basis, these partial electrons are not confined to those edge carbons, but can act collectively when interacting with another radical. So here the "localization" (or "localized" state) at the edge sites is meant to be with respect to the inner sites.

Two pieces of evidence support the idea that zigzag edge π-electrons can respond collectively to an attacking radical. First, we find that after the formation of a C-H bond at an edge coverage of 1/6, the local magnetic moments on the intact carbon atoms on the same edge greatly decrease. In other words, the electronic states at the zigzag edge act together when a C-H bond is formed at one of the edge carbon sites, even though the electron density is distributed evenly at the edge carbon atoms before the bonding. The second piece of evidence is that, the BDE is found to decrease with the edge coverage because there are fewer edge electrons available on a basis of per C-H bond formed when the coverage increases. This is especially the case for a coverage of 1 (Table IV and Fig. 5), where the BDE decreases to 1.93 eV. Although repulsion between H atoms of newly formed, neighboring C-H bonds may also contribute to the decreased BDE for the high coverage of 1, the amount is estimated to small (< 0.02 eV between two neighboring C-H groups).[35]

In addition to H, we also examine other common radicals and list the BDEs in Table V for an edge coverage of 1/6. One can see that like the C-H bond, the C-OH and C-CH$_3$ bonds formed at the edge have a BDE of 50-70% of C$_2$H$_5$-OH and C$_2$H$_5$-CH$_3$ bonds. The structures of the C-OH and C-CH$_3$ bonds that are formed are displayed in Fig. 6. For the halogens, the BDE of edge-X decreases from F to I and follows the same trend as that of C$_2$H$_5$-X. This trend can be attributed to the decreasing electronegativity from F to I. Being the most electronegative, F has the greatest BDE. Moreover, one notes that the BDE ratio of edge-X to C$_2$H$_5$-X increases dramatically from Cl to F, indicating the extraordinary ability of F to pull electrons from the GNR's zigzag edge.

**C. Comparison of ZGNR with armchair-edged GNR, nanotubes, and graphene.**

We have demonstrated the "partial radical" nature of the GNR's zigzag edges. Now we examine how these zigzag edge carbon atoms differ from other sp$^2$-carbon atoms by calculating the BDEs for the C-H bonds formed on a graphene sheet, a metallic nanotube, a semiconducting nanotube, and a graphene nanoribbon with armchair edges (Fig. 7 displays the optimized structures). From Table VI, one can see that the graphene sheet has the lowest reactivity toward a hydrogen radical, whereas the zigzag edges have the highest, and the nanotubes and the armchair edge are in between.

A single graphene sheet is a zero-gap semiconductor (see the LDOS in Fig. 8), and due to its stable π-electron system and flat structure, the interaction between a graphene sheet and an isolated radical ($S=1/2$)[36] has been demonstrated to be weak (usually, BDE < 1 eV).[37,38] The (5,5) carbon nanotube is metallic with low DOS at the Fermi level, whereas the (10,0) carbon nanotube is a small-gap semiconductor (Fig. 8). The two tubes' C-H BDEs are 0.6-0.8 eV higher than that of graphene, mainly due to the tubes' curvature. Our C-H BDEs for the (10,0) and (5,5) tubes agree very well with a previous periodic DFT-GGA study using all-electron Gaussian basis sets.[39] The slightly greater BDE for the (5,5) tube compared to the (10,0) tube may result from its metallic character. The band gap in the armchair-edged GNR (AGNR) depends on the ribbon width.[40] For the AGNR in Fig. 7(d), like the semiconducting (10,0) nanotube, it has a small band gap (Fig. 8), and their C-H BDEs are also similar. The zigzag-edged GNR has a significantly higher C-H BDE than the armchair-edged GNR, nanotubes, or a graphene sheet. By comparing their LDOS (Fig.s 4 and 8), one can see that the ZGNR's unique electronic structure has a substantial peak near the Fermi level, which directly leads to its stronger bonding to hydrogen.

The much stronger bonding of radicals to ZGNRs points to a possibility that ZGNRs may be able to dissociate molecules, such as $H_2$, $CH_3OH$, etc. For example, using data in Table V and the experimental H-H BDE (4.52 eV),[34] one can obtain an energetic change of ~ -1.2 eV for $H_2$ (gas)$\rightarrow$ 2H (adsorbed) at the zigzag edge. Investigation of this kind of dissociation feasibility is planned for the future.

**D. Implications for experimental studies of carbon materials.**

Recent confirmation of the localized electronic state at an HOPG surface's step edge by STS[19] is highly encouraging. The HOPG surface's step edge can be viewed as a semi-infinite graphene sheet, and its reactivity has been demonstrated by electrochemical experiments[41-44] and related to the edge state's electronic structure.[38] However, it would be more exciting to explore the chemical reactivity of ZGNRs as we have predicted in this work. The first challenge for experiments is how to make ZGNRs. We suggest the derivation of ZGNRs from single wall carbon nanotubes (SWCNT), because the latter have been synthesized and separated by many methods and are commercially available. Fig. 9 shows how unzipping a (5,5) SWCNT leads to a ZGNR (N=10). Although how to achieve the transformation in Fig. 9 is by no means clear, it does offer a possible direction to make ZGNRs.

The chemical reactivity that we have shown for the π-electrons of ZGNRs may also contribute to the chemical behaviors of carbon materials such as carbon blacks, carbon fibers, and glassy carbons, which tend to have a large fraction of edge sites. However, due to the structural heterogeneities of these carbon materials, the direct connection between their chemical properties and what we have shown in this work for ZGNRs is hard to establish.

**IV. SUMMARY AND CONCLUSIONS**

First principles density functional theory with the generalized-gradient approximation for electron exchange-correlation was used to study the magnetic, electronic, and chemical properties of graphene nanoribbons with hydrogen-terminated zigzag edges. The results show that the ground state of the zigzag-edged ribbons is antiferromagnetic (AFM) while the ferromagnetic state is only slightly higher in energy. In the AFM phase, the carbon atoms at the edges are found to have a magnetic moment of ~0.14 $\mu_B$, and the local density of states at these carbon atoms shows a strong peak just below the Fermi level for the majority spin and another just above the Fermi level for the minority spin. These features result from the localized electronic state at the zigzag edges and led us to propose a "partial radical" concept to characterize the chemical reactivity of those zigzag edge carbons. By computing the bond dissociation energy (BDE) of the bonds formed between the edge carbon and common radicals, we found that the edge C-X BDE (X=H, OH, $CH_3$, F, Cl, Br, and I) is 40%-80% of the experimental $C(sp^3)$-X BDE, and thus demonstrated the validity of this "partial radical" concept. By comparing the zigzag edge's C-H BDE with those of a graphene sheet, nanotubes, and the armchair edge, we showed that the zigzag edge is indeed unique in that it has the highest BDE, at least 1.2 eV stronger than the others, due to the presence of the edge state near the Fermi level. Recent confirmation of the localized state at the zigzag edge lends further support to our study. We hope that our results will stimulate experimental interest in exploring the unique chemistry of graphene nanomaterials.

**Acknowledgement**.

This work was supported by Office of Basic Energy Sciences, U.S. Department of Energy under Contract No. DE-AC05-00OR22725 with UT-Battelle, LLC, and used resources of the National Center for Computational Sciences at Oak Ridge National Laboratory. DEJ thanks Dr. Chengdu Liang for stimulating discussion about carbon materials. BGS acknowledges research conducted at the Center for Nanophase Materials Sciences, supported by the division of Scientific User Facilities, U.S. Department of Energy.

References and footnotes

[1] H. J. Rader, A. Rouhanipour, A. M. Talarico, V. Palermo, P. Samori, and K. Mullen, Nat. Mater. **5**, 276 (2006).

[2] S. Stankovich, D. A. Dikin, G. H. B. Dommett, K. M. Kohlhaas, E. J. Zimney, E. A. Stach, R. D. Piner, S. T. Nguyen, and R. S. Ruoff, Nature **442**, 282 (2006).

[3] K. Nakada, M. Fujita, G. Dresselhaus, and M. S. Dresselhaus, Phys. Rev. B **54**, 17954 (1996).

[4] K. Kobayashi, Phys. Rev. B **48**, 1757 (1993).

[5] D. J. Klein, Chem. Phys. Lett. **217**, 261 (1994).

[6] D. J. Klein and L. Bytautas, J. Phys. Chem. A **103**, 5196 (1999).

[7] M. Fujita, K. Wakabayashi, K. Nakada, and K. Kusakabe, J. Phys. Soc. Jpn. **65**, 1920 (1996).

[8] M. Fujita, M. Igami, and K. Nakada, J. Phys. Soc. Jpn. **66**, 1864 (1997).

[9] K. Nakada, M. Igami, and M. Fujita, J. Phys. Soc. Jpn. **67**, 2388 (1998).

[10] K. Wakabayashi, M. Fujita, H. Ajiki, and M. Sigrist, Phys. Rev. B **59**, 8271 (1999).

[11] T. Kawai, Y. Miyamoto, O. Sugino, and Y. Koga, Phys. Rev. B **62**, R16349 (2000).

[12] K. Harigaya, Chem. Phys. Lett. **340**, 123 (2001).

[13] K. Wakabayashi and K. Harigaya, J. Phys. Soc. Jpn. **72**, 998 (2003).

[14] A. Yamashiro, Y. Shimoi, K. Harigaya, and K. Wakabayashi, Phys. Rev. B **68**, 193410 (2003).

[15] K. I. Sasaki, S. Murakami, and R. Saito, J. Phys. Soc. Jpn. **75**, 74713 (2006).

[16] K. Sasaki, S. Murakami, and R. Saito, Appl. Phys. Lett. **88**, 113110 (2006).

[17] Z. Klusek, Z. Waqar, E. A. Denisov, T. N. Kompaniets, I. V. Makarenko, A. N. Titkov, and A. S. Bhatti, Appl. Surf. Sci. **161**, 508 (2000).

[18] Y. Kobayashi, K. Fukui, T. Enoki, and K. Kusakabe, Phys. Rev. B **73**, 125415 (2006).

[19] Y. Niimi, T. Matsui, H. Kambara, K. Tagami, M. Tsukada, and H. Fukuyama, Phys. Rev. B **73**, 85421 (2006).

[20] Y. Shibayama, H. Sato, T. Enoki, and M. Endo, Phys. Rev. Lett. **84**, 1744 (2000).

[21] T. Enoki and Y. Kobayashi, J. Mater. Chem. **15**, 3999 (2005).

[22] Y.-W. Son, M. L. Cohen, and S. G. Louie, Nature **444**, 347 (2006).

[23] L. R. Radovic and B. Bockrath, J. Am. Chem. Soc. **127**, 5917 (2005).

[24] H. Lee, Y. W. Son, N. Park, S. W. Han, and J. J. Yu, Phys. Rev. B **72**, 174431 (2005).

[25] G. Kresse and J. Furthmüller, Phys. Rev. B **54**, 11169 (1996).

[26] G. Kresse and J. Furthmüller, Comput. Mater. Sci. **6**, 15 (1996).

[27] J. P. Perdew, K. Burke, and M. Ernzerhof, Phys. Rev. Lett. **77**, 3865 (1996).

[28] P. E. Blöchl, Phys. Rev. B **50**, 17953 (1994).

[29] G. Kresse and D. Joubert, Phys. Rev. B **59**, 1758 (1999).

[30] Here we use the word "ferromagnetic" in a rather loose sense, just to indicate that the spins in the two edges are all up. Strictly speaking, this phase should be called ferrimagnetic because some carbon atoms inbetween the two edges are spin down.

[31] Here N is the ribbon width, and U is the on-site Coulomb repulsion and t is the transfer integral in the Hubbard model.

[32] C. Herring, in *Magnetism*, edited by G. T. Rado and H. Suhl (Academic Press, New York, 1966), Vol. 4.

[33] Zero-point-energy (ZPE) was not included in our calculations of BDEs. Including ZPE corrections was estimated to decrease the calculated BDEs by ~0.2 eV for C-H bonds and by a much less amount for other bonds in the present work.

[34] Y. R. Luo, *Handbook of Bond Dissociation Energies in Organic Compounds*. (CRC Press, Boca Raton, FL., 2002).

[35] The estimate is based on the energy difference between two zigzag systems. Both systems have six edge carbon atoms on each of the two edges with one edge terminated with six hydrogen atoms and the other edge with only two hydrogen atoms. On this other edge, the two hydrogen atoms are separated by two un-hydrogen-terminated carbon atoms in the first system but next to each other in the second system. This energy difference is calculated to be 0.02 eV, which means that the repulsion between neighboring C-H groups should be ~0.02 eV.

[36] Here $S$ is the radical's total spin moment.

[37] X. W. Sha and B. Jackson, Surf. Sci. **496**, 318 (2002).

[38] D. E. Jiang, B. G. Sumpter, and S. Dai, J. Phys. Chem. B **110**, 23628 (2006).

[39] V. Barone, J. Heyd, and G. E. Scuseria, J. Chem. Phys. **120**, 7169 (2004).

[40]Y.-W. Son, M. L. Cohen, and S. G. Louie, Phys. Rev. Lett. **97**, 216803 (2006).

[41]P. Allongue, M. Delamar, B. Desbat, O. Fagebaume, R. Hitmi, J. Pinson, and J. M. Saveant, J. Am. Chem. Soc. **119**, 201 (1997).

[42]K. Ray and R. L. McCreery, Anal. Chem. **69**, 4680 (1997).

[43]J. K. Kariuki and M. T. McDermott, Langmuir **15**, 6534 (1999).

[44]A. H. Holm, R. Moller, K. H. Vase, M. D. Dong, K. Norrman, F. Besenbacher, S. U. Pedersen, and K. Daasbjerg, New J. Chem. **29**, 659 (2005).

TABLE I. Relative energies of different magnetic phases for zigzag-edged graphene nanoribbons.

| Width (N) | $E_{NM}$ (meV/cell)[a] | $E_{FM}$ (meV/cell)[a] | $E_{AFM}$ (meV/cell)[a] |
|---|---|---|---|
| 4 | 0 | −37 | −51 |
| 5 | 0 | −50 | −62 |
| 6 | 0 | −60 | −72 |

[a] The unit cells are shown in FIG. 2.

TABLE II. Total magnetic moments for the ferromagnetic phase ($M_{total-FM}$) and local magnetic moments on the edge carbons for the ferromagnetic and antiferromagnetic phases ($M_{edge-FM}$ and $M_{edge-AFM}$) of zigzag-edged graphene nanoribbons.

| Width (N) | $M_{total-FM}$ ($\mu_B$/cell)[a] | $M_{edge-FM}$ ($\mu_B$) | $M_{edge-AFM}$ ($\mu_B$) |
|---|---|---|---|
| 4 | 0.396 | 0.121 | 0.139 |
| 5 | 0.436 | 0.130 | 0.143 |
| 6 | 0.470 | 0.136 | 0.145 |

[a] The unit cells are shown in FIG. 2.

TABLE III. Bond dissociation energy (BDE) for the bond between a zigzag carbon atom and hydrogen.[a]

| Width (N) | 4 | 5 | 6 |
|---|---|---|---|
| BDE (eV) | 2.82 | 2.86 | 2.87 |

[a] The coverage is at 1/6 H/edge-C (see FIG. 5a).

TABLE IV. Coverage dependence of bond dissociation energy (BDE) of zigzag edge C-H bonds (N=5).

| Coverage (H/edge-C) | 1/6 | 1/3 | 1 |
|---|---|---|---|
| BDE (eV) | 2.86 | 2.73 | 1.93 |

TABLE V. Comparison of bond dissociation energy (BDE, in eV) of zigzag edge-X bonds with experimental BDE of $C_2H_5$-X.[a]

| Radical: X | H | OH | $CH_3$ | F | Cl | Br | I |
|---|---|---|---|---|---|---|---|
| BDE (Edge-X) | 2.86 | 2.76 | 2.22 | 3.71 | 2.18 | 1.65 | 1.18 |
| BDE ($C_2H_5$-X)[b] | 4.358 | 4.055 | 3.838 | 4.904 | 3.651 | 3.036 | 2.420 |

[a] The coverage is at 1/6 X/edge-C (see Fig. 6); zero-point-energy corrections not included.
[b] Experimental values, from Ref. 34. DFT-GGA usually predicts well BDEs for small organic molecules; e.g., our computed BDE for $C_2H_5$-H (4.14 eV) is good agreement with experiment. So here we use experimental BDEs of $C_2H_5$-X for comparison, instead of theoretical ones.

TABLE VI. Comparison of C-H bond dissociation energy (BDE, in eV) for graphene, carbon nanotubes (CNT), and armchair-edged (AGNR) and zigzag-edged (ZGNR) graphene nanoribbon.[a]

| $sp^2$ carbon | Graphene | CNT (10,0) | CNT (5,5) | AGNR | ZGNR |
|---|---|---|---|---|---|
| BDE (C-H) | 0.83 | 1.41 | 1.61 | 1.55 | 2.86 |

[a] See Fig. 7 for the relative coverages of H.

**FIG. 1.** Cutting through an infinite graphene sheet (a) to obtain a semi-infinite sheet with a hydrogen-terminated zigzag (b) or armchair (c) edge. Carbon atoms are shown in black and hydrogen atoms are in gray. The same color scheme is used in FIGs. 2, 3, 5-7.

**FIG. 2.** Zigzag-edged graphene nanoribbons with different width. The rectangular boxes indicate the unit cell. The ribbons are infinitely repeated along the $x$ axis.

**FIG. 3.** Local magnetic moments (in $\mu_B$) for the ferromagnetic (FM) and antiferromagnetic (AFM) phases of a zigzag-edged graphene nanoribbon (N=4).

**FIG. 4.** Local density of states plots for the p-orbital of an edge carbon atom in the nonmagnetic (NM), ferromagnetic (FM), and antiferromagnetic (AFM) phases of a zigzag-edged graphene nanoribbon (N=4). Arrows indicate the direction of spin polarization.

**FIG. 5.** Chemical reaction of hydrogen atoms with a graphene nanoribbon's zigzag edge at different edge coverages: (a) 1/6, (b) 1/3, and (c) 1. Coverage is expressed as reacted hydrogen per edge carbon (N=5).

**FIG. 6.** Reaction of OH (a) and $CH_3$ (b) with a graphene nanoribbon's zigzag edge (N=5). The white ball in (a) is an oxygen atom.

**FIG. 7.** Reaction of a hydrogen atom with (a) a graphene sheet, (b) (10,0) carbon nanotube, (c) (5,5) carbon nanotube, (d) a graphene nanoribbon's armchair edge, and (e) a graphene nanoribbon's zigzag edge.

**FIG. 8.** Local density of states plots for the p-orbital of an carbon atom in a graphene sheet, (10,0) carbon nanotube, (5,5) carbon nanotube, and an armchair edge.

**FIG. 9.** Unzipping a (5,5) single wall carbon nanotube (a) leads to a zigzag-edged graphene nanoribbon (b) with a width of N=10. In (b), left to right indicates the direction of the ribbon's long axis, and the zigzag edges are at the top and the bottom.

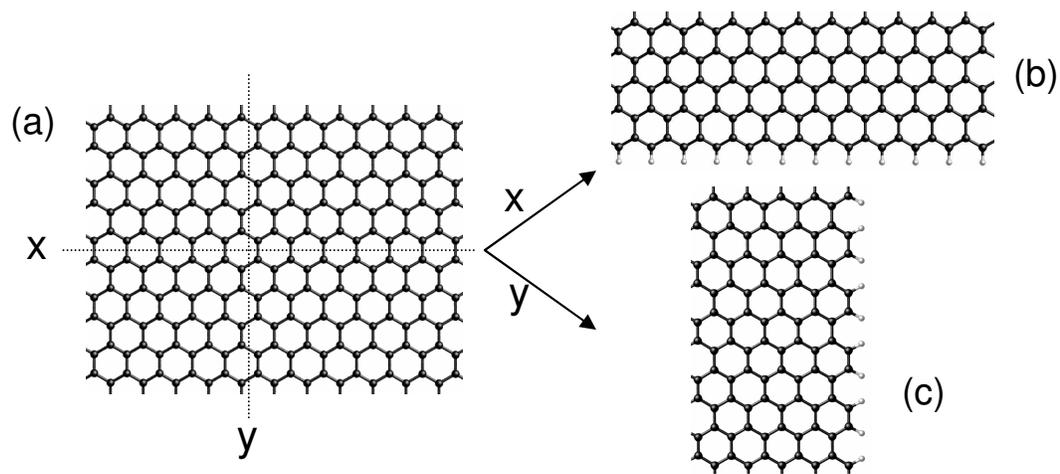

Figure 1, Jiang et al., Journal of Chemical Physics

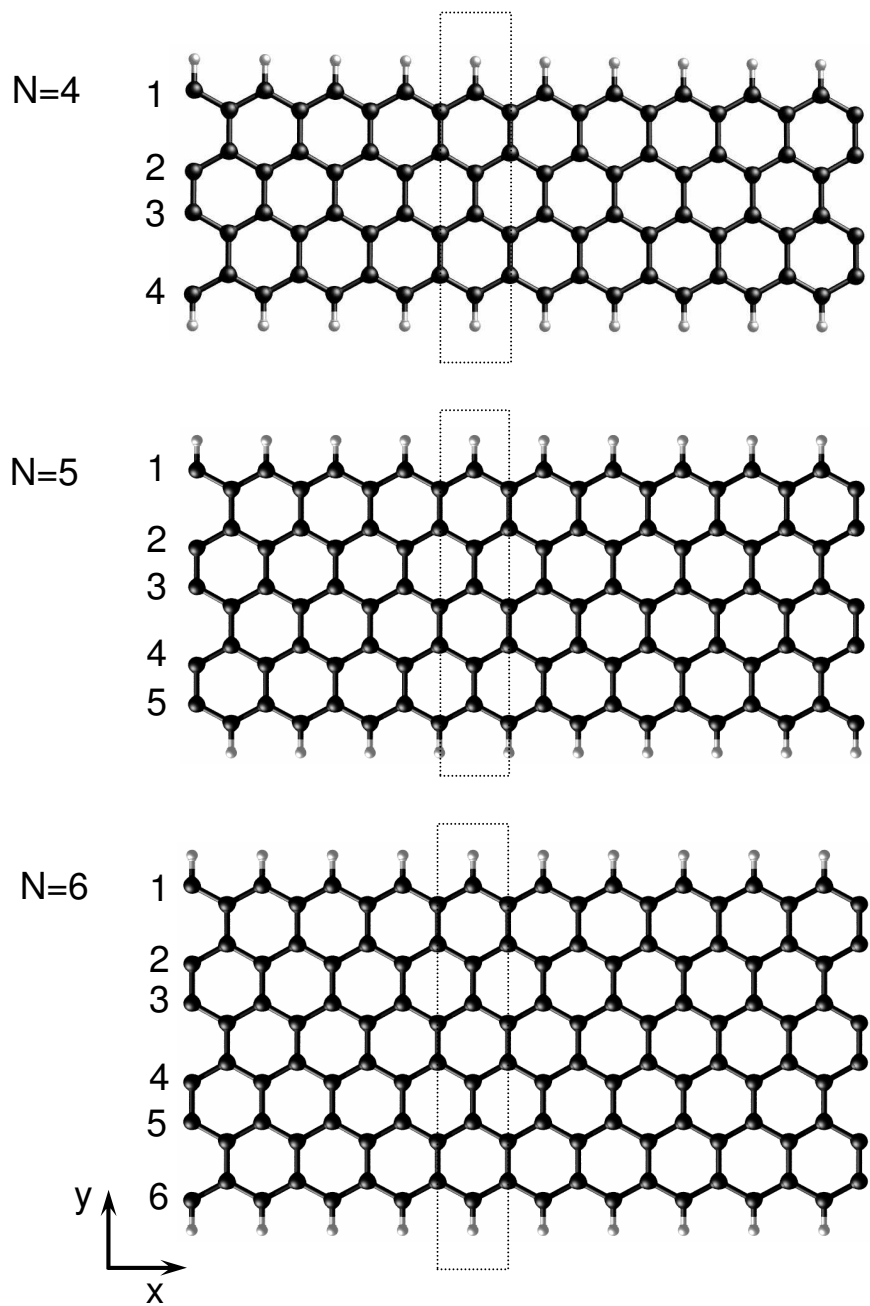

Figure 2, Jiang et al., Journal of Chemical Physics

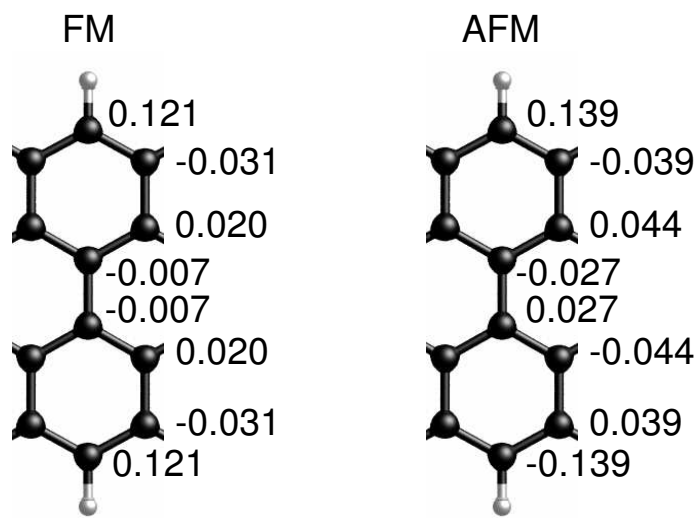

Figure 3, Jiang et al., Journal of Chemical Physics

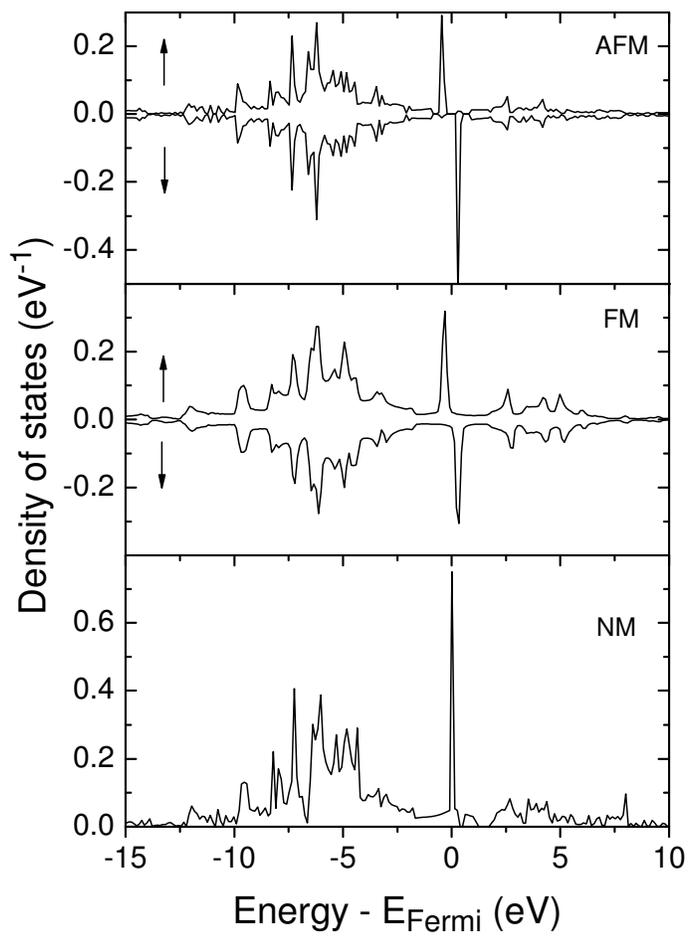

Figure 4, Jiang et al., Journal of Chemical Physics

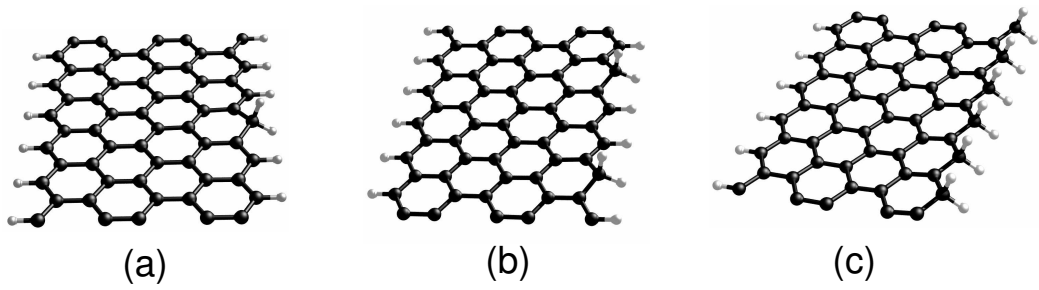

Figure 5, Jiang et al., Journal of Chemical Physics

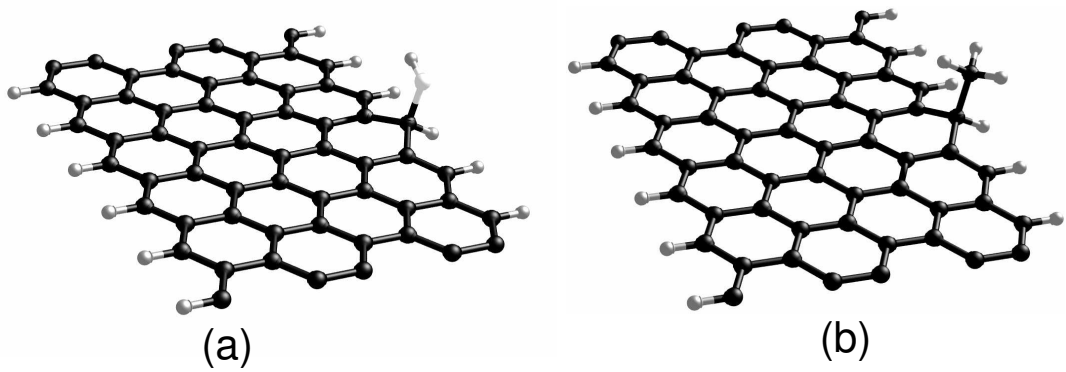

(a)   (b)

Figure 6, Jiang et al., Journal of Chemical Physics

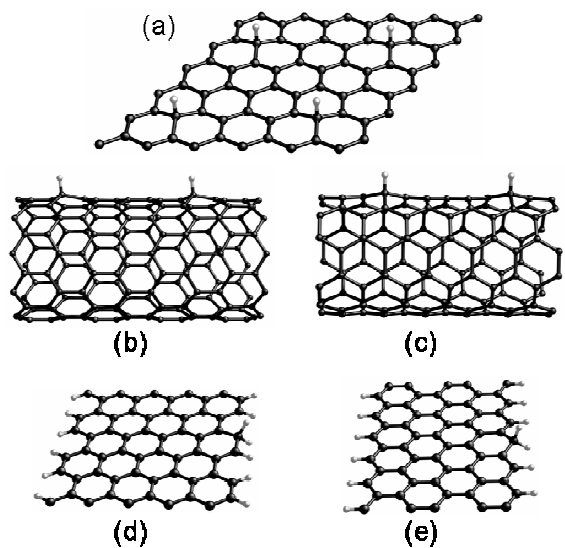

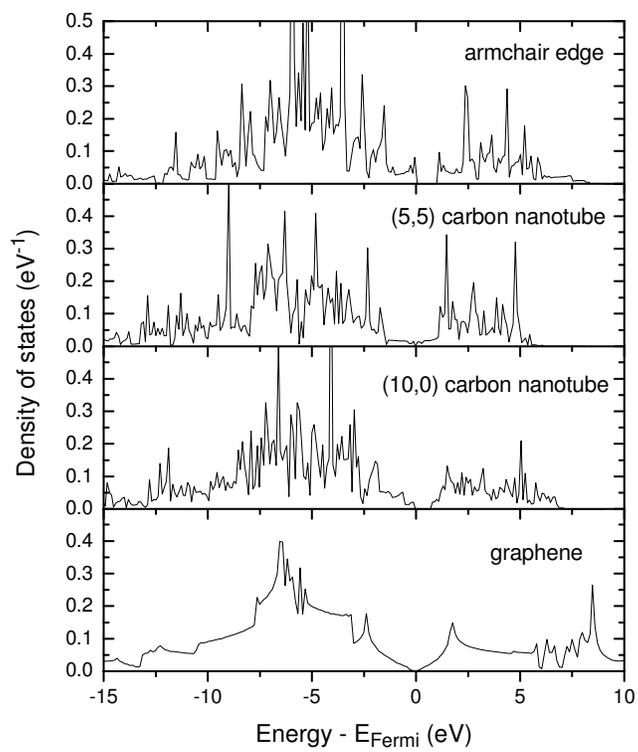

Figure 8, Jiang et al., Journal of Chemical Physics

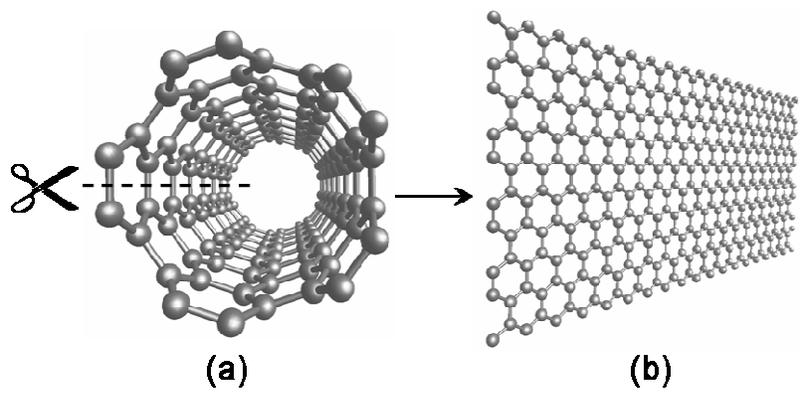

Figure 9, Jiang et al., Journal of Chemical Physics